\documentstyle[12pt]{article}
\def\R{{\rm I\hspace{-.15em}R}}

\def\1{\mbox{I\hspace{-.15em}1}}
\def\C{\hspace{3pt}{\rm l\hspace{-.47em}C}}
\catcode `\@=11 \@addtoreset{equation}{section}
\def\theequation{\arabic{section}.\arabic{equation}}
\def\ba{\begin{array}}
\def\ea{\end{array}}

\def\1{\mathbf {Id} }

\def\b{\begin{equation}}
\def\e{\end{equation}}

\begin{document}

\title{\textbf{Covariant Quantization of ``Massive'' Spin-$\frac{3}{2}$ Fields in the de Sitter Space}}
\author{M. V. Takook$^{1}$
\thanks{E-mail: takook@razi.ac.ir},
A. Azizi$^{1}$, E. Babaian$^{2}$ }

\maketitle

\centerline{\it $^1$ Department of Physics, Razi University,
Kermanshah, IRAN}\centerline{\it $^2$ Department of Physics, Science
and Research Branch,} \centerline{\it Islamic Azad University,
Tehran, IRAN}\vspace{15pt}

\begin{abstract}

We present a covariant quantization of the free ``massive"
spin-$\frac{3}{2}$ fields in four-dimensional de Sitter space-time
based on analyticity in the complexified pseudo-Riemannian manifold.
The field equation is obtained as an eigenvalue equation of the
Casimir operator of the de Sitter group. The solutions are
calculated in terms of coordinate-independent de Sitter plane-waves
in tube domains and the null curvature limit is discussed. We give
the group theoretical content of the field equation. The Wightman
two-point function $S^{i \bar j}_{\alpha\alpha'}(x,x')$ is
calculated. We introduce the spinor-vector field operator
$\Psi_\alpha(f)$ and the Hilbert space structure. A
coordinate-independent formula for the field operator
$\Psi_\alpha(x)$ is also presented.

\end{abstract}

\section{Introduction}

The resent observational data indicated that our universe in first
approximation might be the de Sitter space-time. Quantum field
theory in de Sitter space is a very important subject and it was
studied extensively in the 1980s due to the inflationary model and
linear quantum gravity. This space is the simplest possible
generalization of the Minkowski space-time. The quantization of
various fields (scalar, spinor and vector fields) in the de Sitter
space have been studied by several authors [1-3]
. It has been shown that the massive and massless conformally
coupled scalar fields in the de Sitter space correspond to the
principal and complementary series representation of the de Sitter
group, respectively \cite{BROS1,BR}. The massive vector field in the
de Sitter space has been associated with the principal series,
whereas massless field corresponds to the lowest representation of
the vector discrete series in the de Sitter group \cite{TAK1,GARI}.
The massive and massless spin-2 fields in the de Sitter space have
been also associated with the principal series and the lowest
representation of the rank-2
tensor discrete series of the de Sitter group representation, respectively [8-14]
. The importance of the massless spin-2 field in
the de Sitter space is due to the fact that it plays the central role in
quantum gravity and quantum cosmology.

Supergravity was proposed in $1970$ for describing both gravity and
the other interactions and based upon the principle of general
relativity and quantum mechanics to ``grand unified
schemes"\cite{P}. Supersymmetry is a global symmetry between boson
and fermion and in local form (supergravity) includes gravitational
field. In this framework, the fermionic partner of the gravitational
field is a spin-$\frac{3}{2}$ field, which is called gravitino.
Supersymmetry breaking in supergravity leads to massive gravitinos,
and the gravitino gets mass by the super Higgs mechanism, so massive
spin-$\frac{3}{2}$ fields are essential for understanding the
effective description of supergravity processes in the de Sitter
space \cite{L}. The spin-$\frac{3}{2}$ field in flat space-time was
studied by Rarita-Schwinger \cite{W.R} and recently it has been
considered in \cite{G,H}. The gravitino propagator has been derived
in anti-de Sitter space \cite{A}.

It is instructive to perform a covariant quantization of
spinor-vector field in the de Sitter space. The spinor field in the
de Sitter space has
been treated in many of the papers [21-24]
. In this paper spinor-vector field in the de Sitter space is
considered. For simplicity the following units are used:
$$ c=\hbar=1,\:\:\: \left[x^\alpha/H\right]=1,\:\:\:
\left[M\right]=H,$$ where $c,\hbar$ and $H$ are light velocity,
Planck constant and Hubble parameter respectively. In section 2, we
fix the de Sitter space notations and introduce the two independent
Casimir operators. In section 3, spin-$\frac{3}{2}$ field equation
is obtained as eigenvalue equation of the Casimir operator. The
classification of the unitary irreducible representation of the de
Sitter group in terms of the two parameters $p$ and $q$ is
discussed. In the Minkowskian limit these parameters represent a
spin ($s$) and a mass ($m$) that classify the unitary irreducible
representation of the Poincar\'e group. Then, we derive the de
Sitter field equation from the second order Casimir operator.

Section 4 is devoted to solve the field equation. The solution is
written in terms of a polarization spinor-vector part and a de
Sitter-plane wave
 $$ \Psi_\alpha^i (x)={\cal U}^i_\alpha(x,\xi)(Hx.\xi)^\sigma : i=1,2,3,4 ;\alpha=0,1,2,3,4, $$
where $i$, $\alpha$, $\sigma$ are a spinor index, a space-time index
and a complex number respectively. A five-vector $\xi_\alpha$ is a
future directed null vector in ambient space notation:
$$\xi \in
{\cal C}^+=\{ \xi \in \R^5; \;\;\eta_{\alpha \beta}\xi^\alpha
\xi^\beta=(\xi^0)^2-\vec \xi.\vec \xi-(\xi^4)^2=0,\; \xi^0>0 \}.$$
The solution, $(Hx.\xi)^\sigma $, had been introduced in the context
of harmonic analysis on de Sitter space in the framework of the
$SO(1,4)$ representation theory by Molchanov \cite{MO} and it has
been completely developed by Bros et al. \cite{BROS1,BR}. The plane
wave in curved space, such as three-dimensional Lobachevsky and
Riemann models, were given by some authors \cite{SH}. In contrast to
Minkowski space, ${\cal U}^i_\alpha(x,\xi)$ is a function of the
space-time point $x^\alpha$, because the momentum operators acquire
a spin part \cite{BOG,BO}. The  spinor-vector ${\cal
U}^i_\alpha(x,\xi)$ can be fixed such that in the null curvature
limit one obtain the spinor-vector in the Minkowskian space. These
solutions are not globally defined due to the ambiguity concerning
to the phase factor. For solving this problem the solution is
considered in the complex de Sitter space \cite{BR}. This notation
permits us to define the solution globally on de Sitter hyperboloid
and independent of the choice of the metrics.

In section 5, the Wightman two-point function $S(x,y)$ is
calculated. This function satisfies the conditions of: a)
positivity, b) locality, c) covariance, and d) normal analyticity.
Normal analyticity allows one to define the Wightman two-point
function $S(x,y)$ as the boundary value of the analytic function
${\cal S}(z_1,z_2)$ from the tube domains. The normal analyticity is
related to the Hadamard condition, which selects an unique vacuum
state \cite{BR,T.,BOG}.  ${\cal S}(z_1,z_2)$ is defined in terms of
spinor de Sitter plane-waves in their tube domains. In section
6, we introduce the Hilbert space structure and define the field
operators $\Psi_\alpha (f)$. We also give a coordinate-independent
formula for the field operator $\Psi_\alpha (x)$. Finally, a brief
conclusion and an outlook are given in section 7.

\section{The de Sitter space notations}

The de Sitter space is conveniently seen as a hyperboloid embedded
in a five-dimensional Minkowski space:
$$ X_H=\{x \in \R^5;x^2=\eta_{\alpha\beta} x^\alpha     x^\beta
=-H^{-2}=-\frac{3}{\Lambda}\},\;
\eta_{\alpha\beta}=\mbox{diag}(1,-1,-1,-1,-1),$$ where $\Lambda$ is
a positive cosmological constant. The metric is
$$ds^{2}=\eta_{\alpha\beta}dx^{\alpha}dx^{\beta}\vert_{x^2=-H^{-2}}=g_{\mu\nu}^{dS}dX^{\mu}dX^{\nu},\;\;\mu=0,1,2,3,$$
where the $X^\mu$ is space-time intrinsic coordinates on the de
Sitter hyperboloid. A spinor-tensor field
$\Psi_{\alpha_1..\alpha_l}(x)$ on $X_H$ can be viewed as an
homogeneous function on $ \R^{5}$ with an arbitrary degree of
homogeneity $\sigma$ and the transversality condition \cite{dir}:
$$x\cdot\partial \Psi=\sigma \Psi,$$
$$x\cdot \Psi(x)=0.$$
The tangential or transverse derivative on the de Sitter space is
defined as
\begin{equation}
\partial^\top_\alpha=\theta_{\alpha\beta}\partial^\beta=\partial_\alpha
+H^2x_\alpha x\cdot\partial,\ \ \ \ x\cdot\partial^\top=0,
\end{equation}
where $\theta_{\alpha\beta}=\eta_{\alpha\beta}+H^2x_{\alpha}x_{ \beta}$
is transverse projection tensor ($\theta_{\alpha\beta}\;
x^{\alpha}=\theta_{\alpha\beta} \; x^{\beta}=0$).

The kinematical group of the de Sitter space is the $10$-parameter
group SO$_0(1,4)$ (connected component of the identity), which is
one of the two possible deformations of the Poincar\'e group (the
other one being SO$_0(2,3)$ ). The unitary irreducible
representations of SO$_{0}(1,4)$ are characterized by the
eigenvalues of the two Casimir operators $Q^{(1)}$ and $Q^{(2)}$.
These operators commute with the group generators and they are
constant in each unitary irreducible representation. They read

      \b Q^{(1)}=-\frac{1}{2}L_{\alpha\beta}L^{\alpha\beta}\:\:\:\:,\:\:\:\:Q^{(2)}=-W_\alpha W^\alpha\;\;\;,\;\;\;W_\alpha =\frac{1}{8}
      \epsilon_{\alpha\beta\gamma\delta\eta} L^{\beta\gamma}L^{\delta\eta},\e
where $\epsilon_{\alpha\beta\gamma\delta\eta}$ is the usual
antisymmetrical tensor in $\R^5$ and
$L_{\alpha\beta}=M_{\alpha\beta}+S_{\alpha\beta}$ is an
infinitesimal generator. The orbital part $M_{\alpha\beta}$ is
          \b  M_{\alpha \beta}=-i(x_\alpha \partial_\beta-x_\beta
      \partial_\alpha)=-i(x_\alpha\partial^\top_\beta-x_\beta
        \partial^\top_\alpha).\e
In order to precise the action of the spinorial  part
$S_{\alpha\beta}$ on a tensor field or spinor-tensor field, one must
treat separately the integer and half-integer cases. Integer spin
fields can be represented by tensor fields of rank $l$,
 $\Psi_{\gamma_1...\gamma_l}(x)$, and the spinorial action reads
\cite{GAZ}
       \b S_{\alpha \beta}^{(l)}\Psi_{\gamma_1......\gamma_l}=-i\sum^l_{i=1}
          \left(\eta_{\alpha\gamma_i}
        \Psi_{\gamma_1....(\gamma_i\rightarrow\beta).... \gamma_l}-\eta_{\beta\gamma_i}
          \Psi_{\gamma_1....(\gamma_i\rightarrow \alpha).... \gamma_l}\right),\e
where $(\gamma_i\rightarrow\beta)$ means $\gamma_i$ index replaced
with $\beta$. Half-integer spin fields with spin $s=l+\frac{1}{2}$
are represented by four component spinor-tensor $\Psi_{
\gamma_{1}..\gamma_{l} }^{i}$ with $i=1,2,3,4$. In this case, the
spinorial part is
$$S_{\alpha\beta}^{(s)}=S_{\alpha\beta}^{(l)}+S_{\alpha\beta}^{(\frac{1}{2})}\qquad
\mbox{with}\qquad
S_{\alpha\beta}^{(\frac{1}{2})}=-\frac{i}{4}\left[\gamma_{\alpha},\gamma_{\beta}\right].$$
The five matrices $\gamma_{\alpha}$ are determined by the relations
\cite{TAK2,BOG,TAK3}
$$\gamma^{\alpha}\gamma^{\beta}+\gamma^{\beta}\gamma^{\alpha}
=2\eta^{\alpha\beta}\qquad
\gamma^{\alpha\dagger}=\gamma^{0}\gamma^{\alpha}\gamma^{0}\;,$$
$$ \gamma^0=\left( \begin{array}{clcr} I & \;\;0 \\ 0 &-I \\ \end{array} \right)
      ,\gamma^4=\left( \begin{array}{clcr} 0 & I \\ -I &0 \\ \end{array} \right) , $$ \b
   \gamma^1=\left( \begin{array}{clcr} 0 & i\sigma^1 \\ i\sigma^1 &0 \\
    \end{array} \right)
   ,\gamma^2=\left( \begin{array}{clcr} 0 & -i\sigma^2 \\ -i\sigma^2 &0 \\
      \end{array} \right)
   , \gamma^3=\left( \begin{array}{clcr} 0 & i\sigma^3 \\ i\sigma^3 &0 \\
      \end{array} \right),\e
where $\sigma_i$ are Pauli matrices and $I$ is a $2\times2$ unit
matrix. The Casimir operators are simple to manipulate in ambient
space notation. Since $Q^{(1)}$ is a second order derivative
operator, it is used for obtaining the field equation in this paper.
In particular, it is easy to show that for a $l$-rank tensor field
$\Psi_{\gamma_{1}...\gamma_{l}}(x)$ one has
 \b Q_l^{(1)}\Psi=Q_0^{(1)}\Psi-2\Sigma_1 \partial x .\Psi+2\Sigma_1 x \partial.
           \Psi+2\Sigma_2 \eta \Psi'-l(l+1)\Psi,\e
where
   \b Q_l^{(1)}=-\frac{1}{2}L_{\alpha \beta}^{(l)}L^{\alpha \beta(l)}
         =-\frac{1}{2}M_{\alpha \beta}M^{\alpha \beta}-\frac{1}{2} S_{\alpha
           \beta}^{(l)}S^{\alpha
            \beta(l)} -M_{\alpha \beta}S^{\alpha \beta(l)},\e

\b M_{\alpha \beta}S^{\alpha \beta(l)}\Psi(x)= 2\Sigma_1
\partial x
       .\Psi-2\Sigma_1 x \partial.\Psi-2l\Psi.\e
       \b\frac{1}{2}S_{\alpha \beta}^{(l)}S^{\alpha \beta(l)}\Psi=l(l+3)\Psi-
          2\Sigma_2 \eta \Psi',\e

\b Q_0^{(1)}=-\frac{1}{2}M_{\alpha \beta}M^{\alpha \beta}.\e $\Psi'$
is the trace of the $l$-rank tensor $\Psi(x)$ viewed as a
homogeneous function of the variables $x^{\alpha}$ and $\Sigma_p$ is
the non-normalized symmetrization operator:
         \b \Psi'_{\alpha_1...\alpha_{l-2}}=\eta^{\alpha_{l-1}\alpha_l}
          \Psi_{\alpha_1...\alpha_{l-2} \alpha_{l-1}\alpha_l},\e
         \b (\Sigma_p AB)_{\alpha_1...\alpha_l}=\sum_{i_1<i_2<...<i_p}
          A_{\alpha_{i_1}\alpha_{i_2}...\alpha_{i_p}}
          B_{\alpha_1...\not\alpha_{i_1}...\not\alpha_{i_2} ...\not\alpha_{i_p}...\alpha_l}.\e

For half-integer spin fields with spin $s=l+\frac{1}{2}$, the
$S_{\alpha\beta}^{(\frac{1}{2})}$ acts only upon the index i, and we
have \cite{BOG,LESM}
$${\cal S}^{(\frac{1}{2})}_{\alpha\beta}{\cal S}^{\alpha\beta(l)}\Psi(x)=l \Psi(x)-\Sigma_1\gamma(\gamma\cdot
\Psi(x)).$$ In this case, the Casimir operator is
          $$ Q^{(1)}_s=-\frac{1}{2}\left(M_{\alpha \beta}+S_{\alpha \beta}^{(l)}+
                S_{\alpha\beta}^{(\frac{1}{2})}\right)
      \left(M^{\alpha \beta}+S^{\alpha \beta(l)}+S^{\alpha \beta(\frac{1}{2})}\right)$$
   \b =Q^{(1)}_l-\frac{5}{2}+\frac{i}{2}\gamma_{\alpha}\gamma_{\beta}M^{\alpha
\beta}-
             S_{\alpha \beta}^{(\frac{1}{2})}S^{\alpha \beta(l)}.\e
Then we obtain
       \b Q^{(1)}_s\Psi(x)=\left(Q^{(1)}_l-l-\frac{5}{2}+\frac{i}{2}\gamma_{\alpha}\gamma_{\beta}M^{\alpha
       \beta}\right)\Psi(x)+ \Sigma_1 \gamma (\gamma.\Psi(x)) ,\e
or
      $$ Q^{(1)}_s\Psi(x)=\left(-\frac{1}{2}M_{\alpha \beta}M^{\alpha
      \beta}+\frac{i}{2} \gamma_{\alpha}\gamma_{\beta}
       M^{\alpha \beta}-l(l+2)-\frac{5}{2}\right)\Psi(x)$$ \b -2\Sigma_1
         \partial x .\Psi(x)+2\Sigma_1 x \partial.\Psi(x)+2\Sigma_2 \eta
       \Psi'(x)+\Sigma_1 \gamma (\gamma.\Psi(x)).\e
The spin-$\frac{3}{2}$ field equation can be written in terms of
the Casimir operator $Q_{}^{(1)}$, which will be done in the next
section.

\section{ Field equation}

The operator $Q_{\frac{3}{2}}^{(1)}$ commutes with the group
generators and consequently it is constant on each unitary
irreducible representation. In fact, the spinor-vector unitary
irreducible representations can be classified by using the
eigenvalues of $Q^{(1)}$ and the field equation can be written as
\b\left(Q_{\frac{3}{2}}^{(1)}-<Q_{\frac{3}{2}}^{(1)}>\right)\Psi(x)=0.
\e

From Takahashi \cite{TAKA} and Dixmier \cite{DIX}, we get a general
classification scheme for all the unitary irreducible
representations of the de Sitter group, which may be labeled by a
pair of parameters $(p,q)$ with $2p \in N$ and $q \in C$. In terms
of which the eigenvalues of $Q^{(1)}$ and $Q^{(2)}$ are expressed as
follows:
\begin{equation}
<Q^{(1)}>=[-p(p+1)-(q+1)(q-2)], \quad
<Q^{(2)}>=[-p(p+1)q(q-1)]. \label{casimirs}
\end{equation}
For spin-$\frac{3}{2}$ field, according to the possible values of
$p$ and $q$, two types of unitary irreducible representation are
distinguished for the de Sitter group $SO(1,4)$ namely, the
principal and the discrete series. The flat limit indicates that for
the principal series the value of p has the meaning of spin. For the
discrete series case, the only representation which has a physically
meaningful Minkowskian counterpart is $p=q=s$ case. Mathematical
details of the group contraction and the relationship between the de
Sitter and the Poincar\'e groups can be found in \cite{LE,BAC}. The
spin-$\frac{3}{2}$ field representations relevant to the present
work are as follows:
\begin{itemize}
\item[i)] The unitary irreducible representations $U^{\frac{3}{2},\nu}$ in the principal series where
$p = s = \frac{3}{2}$ and $q = \frac{1}{2} + i\nu$ correspond to
the Casimir spectral values:\begin{equation}
<Q_\frac{3}{2}^{(1)}>=\nu^2-\frac{3}{2},\;\;\;\;\nu \in \R \:\:\:
\nu>\frac{3}{2}.
\end{equation}
Note that $U^{\frac{3}{2},\nu}$ and $U^{\frac{3}{2},-\nu}$ are
equivalent.
\item[ ii)]  The unitary irreducible representations $\Pi^{\pm}_{\frac{3}{2},q}$ of the discrete
series, where $p = s = \frac{3}{2}$, correspond to  $$
<Q_\frac{3}{2}^{(1)}>= -\frac{5}{2} ,\; \;\;\;
q=\frac{3}{2},\:\:\:\:\:\Pi^{\pm}_{\frac{3}{2},\frac{3}{2}},
$$
\b <Q_\frac{3}{2}^{(1)}>= -\frac{3}{2} ,\; \;\;\;
q=\frac{1}{2},\:\:\:\:\:\Pi^{\pm}_{\frac{3}{2},\frac{1}{2}}.\e
\end{itemize}

Let us recall at this point the physical content of the principal
series representation from the point of view of a Minkowskian
observer at the limit $H=0$. The principal series unitary
irreducible representation $U^{\frac{3}{2},\nu},\;\;\nu >
\frac{3}{2}$, contracts toward the direct sum of two massive
spinor-vector unitary irreducible representations of the Poincar\'e
group ${\cal P}^<(m,\frac{3}{2})$ and ${\cal P}^>(m,\frac{3}{2})$,
with negative and positive energies, respectively:
$$
U^{\frac{3}{2},\nu}  {H\rightarrow 0 \over \nu \rightarrow \infty }
\longrightarrow {\cal P}^<(m,\frac{3}{2})\bigoplus {\cal
P}^>(m,\frac{3}{2}).
$$
The contraction limit has to be understood through the constraint
$m_H= H\nu$. The quantity $m_H$, supposed to depend on $H$, goes to
the Minkowskian mass $m$ when the curvature goes to zero.

The spin-$\frac{3}{2}$ field in discrete series corresponds to
$\Pi^{\pm}_{\frac{3}{2},\frac{3}{2}}$ and
$\Pi^{\pm}_{\frac{3}{2},\frac{1}{2}}$, in which the sign ${\pm}$
stands for the helicity. In these cases, the two
representations$\Pi^{\pm}_{\frac{3}{2},\frac{3}{2}}$,  with $p = q
=\frac{3}{2}$, have a Minkowskian interpretation. The representation
$\Pi^{\pm}_{\frac{3}{2},\frac{3}{2}}$ has a unique extension to a
direct sum of two unitary irreducible representations $
C(\frac{5}{2}; \frac{3}{2}, 0)$ and $C(-\frac{5}{2}; \frac{3}{2},0)$
of the conformal group $SO(2,4)$ with positive and negative energies
respectively \cite{LE,BA}. The latter restricts to the massless
unitary irreducible representations of Poincar\'e group ${\cal
P}^{>}(0,\frac{3}{2})$ and ${\cal P}^{<}(0,\frac{3}{2})$ with
positive and negative energies respectively and positive helicity.
The following diagrams illustrate these relations: \b \left.
\begin{array}{ccccccc} &             & {\cal C}(\frac{5}{2}, \frac{3}{2},0)
& &{\cal C}(\frac{5}{2}, \frac{3}{2},0)   &\hookleftarrow &{\cal P}^{>}(0,\frac{3}{2})\\
\Pi^+_{\frac{3}{2},\frac{3}{2}} &\hookrightarrow  & \oplus
&\stackrel{H=0}{\longrightarrow} & \oplus  & &\oplus  \\
 &             & {\cal C}(-\frac{5}{2}, \frac{3}{2},0)&
& {\cal C}(-\frac{5}{2}, \frac{3}{2},0)  &\hookleftarrow &{\cal P}^{<}(0,\frac{3}{2}),\\
\end{array} \right. \e
\b \left. \begin{array}{ccccccc}
 &             & {\cal C}(\frac{5}{2}, 0,\frac{3}{2})
& &{\cal C}(\frac{5}{2}, 0,\frac{3}{2}) &\hookleftarrow &{\cal P}^{>}(0,-\frac{3}{2})\\
\Pi^-_{\frac{3}{2},\frac{3}{2}} &\hookrightarrow  & \oplus
&\stackrel{H=0}{\longrightarrow} &  \oplus & &\oplus  \\
&             & {\cal C}(-\frac{5}{2}, 0,\frac{3}{2})&
& {\cal C}(-\frac{5}{2}, 0,\frac{3}{2})   &\hookleftarrow &{\cal       P}^{<}(0,-\frac{3}{2}),\\
\end{array} \right. \e
where the arrows $\hookrightarrow $ designate unique extension. It
is important to note that the representations
$\Pi^{\pm}_{\frac{3}{2},\frac{1}{2}}$ do not have a corresponding
flat limit. \newline Now let us consider the unitary irreducible
representations of the principal series. For spin-$\frac{3}{2}$
field, we obtain the following field equation:
\begin{equation}
\left[Q_\frac{3}{2}^{(1)}-\left(\nu^{2}-\frac{3}{2}\right)\right]\Psi(x)=0,
\label{3.7}
\end{equation} where
$$ Q^{(1)}_{\frac{3}{2}}\Psi(x)=\left(-\frac{1}{2}M_{\alpha \beta}M^{\alpha
      \beta}+\frac{i}{2} \gamma_{\alpha}\gamma_{\beta}
       M^{\alpha \beta}-3-\frac{5}{2}\right)\Psi(x)$$ $$ -2\partial x.\Psi(x)+2x\partial.\Psi(x)+\gamma
       (\gamma.\Psi(x)).$$
If the spin-$\frac{3}{2}$ field satisfies the following subsidiary
conditions:
\begin{itemize}
\item[i)] transversality, $ x.\Psi(x)=0,$
\item[ii)] divergencelessness, $ \partial^\top .\Psi(x)=0,$
\item[iii)] and $\gamma .\Psi(x)=0,$
\end{itemize}
it can be associated with an unitary irreducible representation of
the de Sitter group. Then the action of $Q_\frac{3}{2}^{(1)}$on
$\Psi(x)$ gives
 \b Q^{(1)}_{\frac{3}{2}}\Psi(x)=\left(-\frac{1}{2}M_{\alpha \beta}M^{\alpha
      \beta}+\frac{i}{2} \gamma_{\alpha}\gamma_{\beta}
       M^{\alpha \beta}-\frac{5}{2}-3\right)\Psi(x). \label{3.8}\e
Now by using (\ref{3.7}) and (\ref{3.8}) we obtain first order
spin-$\frac{3}{2}$ field equation such as \cite{TAK2,BOG,TAK3} \b
\left(-i\not
x\not\partial^\top+2i+\nu\right)\Psi_\alpha(x)=0,\label{3.9}\e where
$\not x=\gamma_\alpha x^\alpha$ and $\not\partial^\top=\gamma^\alpha
\partial^\top_\alpha$ . This equation is exactly the same as the de
Sitter-Dirac spin-$\frac{1}{2}$ field equation. This procedure is
similar to obtain the Dirac equation from the Klein-Fock-Gordon
equation. It reduces to the usual Rarita-Schwinger equation in the
Minkowski space-time in the null curvature limit and mass can be
find by
$$\lim_{H\rightarrow 0,\nu\rightarrow \infty} {H\nu}=m. $$

Similar to the spin-$\frac{1}{2}$ field in the de Sitter space
\cite{BOG}, due to the orthogonality of the solutions and obtaining
the Minkowskian solution in the null curvature limit, the adjoint
spinor field ${\overline \Psi}_\alpha(x)$ in ambient space notation
is defined as follows: $${\overline \Psi}_\alpha(x)\equiv
\Psi^{\dag}_\alpha(x){\gamma^0}{\gamma^4}.$$ It satisfies the
following field equation: \b
\overline{\Psi}_\alpha(x)\left[{\gamma^4}\overleftarrow{Q}_{\frac{3}{2}}^{(1)}{\gamma^4}-\left(\nu^2-\frac{3}{2}\right)\right]=0,\label{3.10}
\e
 or equivalently \b {\overline
\Psi}_\alpha(x){\gamma^4}\left(i\overleftarrow{\not\partial}^\top\not x
-2i+\nu\right){\gamma^4}=0. \e The derivative acts to the left in the
usual notation. In the next section the second order field equation $(\ref{3.7})$ will be solved and in the appendix we solve the first order equation $(\ref{3.9})$.

\section{The de Sitter spin-$\frac{3}{2}$ plane waves}

By using the de Sitter plane waves, which were presented by Bros et
al. \cite{BROS1}, we calculated the de Sitter-Dirac plane wave for
spinor field \cite{BOG}. The spinor-vector solution can be written
in terms of the spinor fields (for simplicity from now on we set $H
= 1$):
 \b \Psi_\alpha (x)= Z^\top_\alpha \psi_1+D_{\frac{3}{2}\alpha} \psi_2+\gamma^\top_\alpha\psi_3, \label{4.1}\e
where $\psi_1 ,\psi_2$, and $\psi_3$ are spinor fields.
$\gamma^\top$ is the transverse projection of $\gamma ,
(\gamma^\top_\alpha=\theta_{\alpha \beta} \gamma^\beta)$ and
$D_{\frac{3}{2}\alpha}= -
\partial^\top_\alpha- \gamma^\top_\alpha \not x$. $Z$ is an arbitrary five-component constant vector field: $$
Z^\top_\alpha =\theta_{\alpha\beta} Z^\beta=Z_\alpha+x_\alpha x\cdot
Z,\;x\cdot Z^\top=0.$$ Putting $\Psi_\alpha$ in equation (\ref{3.7})
and using the following identities: \b
Q_{\frac{3}{2}}D_{\frac{3}{2}}=D_{\frac{3}{2}}Q_{\frac{1}{2}},
\label{4.2}\e \b Q_{\frac{3}{2}}\gamma^\top\psi=\gamma^\top
(Q_0-\frac{5}{2}+\not x \not  \partial^\top )\psi(x), \label{4.3}\e
\b \not x \not  \partial^\top  Z^\top \psi=Z^\top \not x \not
\partial^\top \psi- \gamma^\top \not x(Z.x)\psi+x_\alpha\not
x\not z\psi,\label{4.4}\e
 \b Q_{\frac{3}{2}} Z^\top \psi= Z^\top(Q_{\frac{1}{2}}-3)\psi-
 2D_1(Z.x)\psi-\gamma^\top \not x(Z . x)\psi+\gamma^\top (Z . \gamma^\top)\psi,\label{4.5}\e
we find that the spinor fields $\psi_1,\psi_2$, and $\psi_3$ must
obey the following equations:
 \b ( Q_{\frac{1}{2}}-(\nu^2+\frac{3}{2}))\psi_1=0\;\;\mbox{or} \;\;(-i\not x\gamma.\bar{\partial}
+2i+\nu)\psi_1(x)=0,\e \b (
Q_{\frac{1}{2}}-(\nu^2-\frac{3}{2}))\psi_2-2(x.Z)\psi_1=0,\e \b
(Q_0-\frac{5}{2}+\not x \not \bar
\partial-(\nu^2-\frac{3}{2}))\psi_3-3\not x(x.Z)\psi_1+(Z.\gamma^\top) \psi_1=0.\e
It is clear that $\psi_1$ is a ``massive'' spinor field associated
to the principal series $U^{\frac{1}{2},\nu}$ \cite{BOG}. The
divergencelessness condition, $ \partial^\top .\Psi(x)=0$, results
in \b (Z^\top.\partial+4Z.x)\psi_1+(Q_0+\not x \not
\partial^\top )\psi_2+(4\not x+\gamma^\top\cdot \partial^\top
)\psi_3=0.\label{4.9}\e The subsidiary condition $\gamma .\Psi(x)=0$
gives \b \psi_3=-\frac{1}{4}\left(\gamma. Z^\top \psi_1-(\gamma.
\partial^\top +4\not x)\psi_2\right).\label{4.10}\e By using equations $(\ref{4.9})$ and $(\ref{4.10})$, the spinor field $\psi_2$ and $\psi_3$ can be written in terms of spinor field $\psi_1$: $$
\psi_2=\frac{1}{(\nu^2+1)}\left(2x.Z+\frac{2}{3}Z. \partial^\top
+\frac{1}{3}(i\nu+1)\not Z^\top\not x
 \right)\psi_1,$$
 \b \psi_3=-\frac{1}{4}\left[\gamma. Z^\top -\frac{\gamma.
\partial^\top +4\not x}{(\nu^2+1)}\left(2x.Z+\frac{2}{3}Z. \partial^\top
+\frac{1}{3}(i\nu+1)\not Z^\top\not x
 \right)\right]\psi_1. \label{4.11}\e
Then the spinor-vector solution is obtained in the following form:
  $$ \Psi_\alpha(x)=\left(
{Z^\top}_\alpha-\frac{1}{4}\gamma^\top_\alpha\not
Z^\top\right)\psi_1-\left( \partial^\top
_\alpha-\frac{1}{4}\gamma^\top_\alpha\not \partial^\top
\right)\psi_2$$ \b
=\left[{Z^\top}_\alpha-\frac{1}{4}\gamma^\top_\alpha\not Z^\top
-\frac{\partial^\top _\alpha-\frac{1}{4}\gamma^\top_\alpha\not
\partial^\top }{(\nu^2+1)}\left(2x.Z+\frac{2}{3}Z.
\partial^\top +\frac{1}{3}(i\nu+1)\not Z^\top\not x
 \right)\right]\psi_1.\e
This solution can be written in the following compact form: \b
\Psi_\alpha(x)={\cal D}_\alpha\psi_1,\e where
$$ {\cal D}_\alpha=Z^\top_\alpha-\frac{1}{4}\gamma^\top_\alpha\not Z^\top-\frac{1}{\nu^2+1}\left( -\frac{1}{4}(1-i\nu)\gamma^\top_\alpha\not Z^\top+2\partial^\top _\alpha x.Z \right.$$ $$\left. +\frac{2}{3}\partial^\top _\alpha Z. \partial^\top +\frac{1}{3}(i\nu+1)\partial^\top _\alpha \not Z^\top \not x-\frac{1}{3}(i\nu+1)\gamma^\top_\alpha Z.x \not x-\frac{2}{3}\gamma^\top_\alpha Z.x \not \partial^\top  \right.  $$ $$ \left.   -\frac{1}{6}\gamma_\alpha^\top Z. \partial^\top  \not \partial^\top -\frac{1}{6}i\nu\gamma^\top_\alpha \not x Z. \partial^\top-\frac{1}{12}(i\nu+1)\gamma^\top_\alpha\not Z^\top \not x \not \partial^\top\right),
$$ and $\psi_1$ is the solution of de Sitter-Dirac field equation. In the previous paper, the spinor field $\psi_1$ was explicitly calculated and the solutions are given by \cite{TAK2,BOG}  $$(\psi_1)_1=(x.\xi )^{-2+ i \nu}{\cal
V}(x,\xi),$$ \b (\psi_1)_2=(x.\xi)^{-2- i \nu}{\cal U}(\xi),\e where
${\cal V}(x,\xi)=\not x \not \xi{\cal V}(\xi)$ and $\xi \in {\cal
C}^+$. The two spinors ${\cal V}(\xi)$ and ${\cal U}(\xi)$ are \b
{\cal U}^a(\xi)=\frac{\xi^0-\vec \xi. \vec \gamma
 \gamma^0+1}{\sqrt{2(\xi^0+1)}}{\cal U}^a(\stackrel{o}{\xi}_+),\;\;\; {\cal V}^a(\xi)=\frac{1}{\sqrt{2(\xi^0+1)}}{\cal U}^a(\stackrel{o}{\xi}_-), \;\;a=1,2, \e where
  \b {\cal U}^1(\stackrel{o}{\xi}_+) =\frac{1}{\sqrt{2}}\left( \begin{array}{clcr} \alpha\\ \alpha \\ \end{array} \right) ,\; {\cal U}^2             (\stackrel{o}{\xi}_+)=\frac{1}{\sqrt{2}}\left( \begin{array}{clcr} \beta\\ \beta \\ \end{array} \right),\label{4.16}\e
\b {\cal U}^1(\stackrel{o}{\xi}_-) =\frac{1}{\sqrt{2}}\left(
\begin{array}{clcr} \alpha\\-\alpha \\ \end{array} \right) ,\; {\cal
U}^2             (\stackrel{o}{\xi}_-)=\frac{1}{\sqrt{2}}\left(
\begin{array}{clcr} \beta\\-\beta \\ \end{array} \right),\label{4.17}\e with
$\alpha=\left( \begin{array}{clcr}1 \\ 0 \\ \end{array} \right)$ ,
$\beta=\left( \begin{array}{clcr} 0 \\ 1 \\ \end{array} \right)$ and
$ \xi =\stackrel{o}{\xi}_\pm \equiv (1,\vec 0,\pm 1)$.

Finally the two possible solutions for $\Psi_\alpha(x)$ are
$$\Psi^a_{1\alpha}(x)=\left[{Z^\top}_\alpha-\frac{1}{4}\gamma^\top_\alpha\not
Z^\top -\frac{ \partial^\top
_\alpha-\frac{1}{4}\gamma^\top_\alpha\not
\partial^\top}{\nu^2+1}\left(2x.Z+\frac{2}{3}Z. \partial^\top
+\frac{1}{3}(i\nu+1)\not Z^\top\not x \right)\right]$$\b\not x
\not\xi {\cal V}^a(\xi)(x.\xi )^{-2+ i \nu} \equiv{\cal
V_\alpha}^a(x,\xi,Z)(x.\xi)^{-2+ i \nu},\label{4.19}\e and
$$\Psi^a_{2\alpha}(x)=\left[{Z^\top}_\alpha-\frac{1}{4}\gamma^\top_\alpha\not
Z^\top -\frac{ \partial^\top
_\alpha-\frac{1}{4}\gamma^\top_\alpha\not
\partial^\top}{\nu^2+1}\left(2x.Z+\frac{2}{3}Z. \partial^\top
+\frac{1}{3}(i\nu+1)\not Z^\top\not x \right)\right]$$ \b{\cal
U}^a(\xi)(x.\xi )^{-2- i \nu}\equiv{\cal
U_\alpha}^a(x,\xi,Z)(x.\xi)^{-2-i\nu},\label{4.20}\e where $a$ is
two spinor states equation (\ref{4.16}) or (\ref{4.17}). In the
following we see that the sign of $\nu$ in the plane wave play the
role of the sign of energy in the Minkowskian limit. By taking the
derivative of the plane wave $(x.\xi)^\sigma,$ the explicit forms of
${\cal U}_\alpha$ and ${\cal V}_\alpha$ are obtained in terms of
$\xi$ [appendix A.2].

The arbitrary constant vector $Z$, which defines the polarization
states in de Sitter space, is fixed in the null curvature limit
$H=0$ (in this part we add $H$). In this limit, $(Hx\cdot
\xi)^{-2-i\nu}$ becomes the plane wave $e^{ ik\cdot X}$ and ${\cal
V}(\xi)$ and ${\cal U}(\xi)$ become the spinors ${\cal V}(k)$ and
${\cal U}(k)$ in the Minkowski space \cite{BOG}, and ${\cal
D}_\alpha$ becomes the vector polarization
$\varepsilon_\mu^{(\lambda)}(k)$ in the Minkowskian space-time
\cite{TAK1}:
$$ \lim_{H \rightarrow 0,\nu\rightarrow \infty}{\cal
U}_{\alpha}(x,\xi,Z)\left[Hx(X)\cdot \xi\right]^{-2-i\nu}\equiv \varepsilon_\mu^{(\lambda)}(k){\cal U
}(k)e^{- ik\cdot X},$$
$$ \lim_{H \rightarrow 0,\nu\rightarrow \infty} {\cal
V}_{\alpha}(x,\xi,Z)\left[Hx(X)\cdot \xi\right]^{-2+i\nu} \equiv
\varepsilon_\mu^{(\lambda)*}(k){\cal V}(k)e^{ ik\cdot X},$$ where
$\lambda$ is taking the three values for three polarization states
of a massive vector field \cite{TAK1}, and $\xi$ is parameterized in
terms of the four-momentum $k$: \b
\xi=\left(\frac{k^0}{mc}=\sqrt{\frac{\vec k^2}{m^2c^2}+1},\frac{\vec
k}{mc},-1\right).\e The four-vector
$\varepsilon^{(\lambda)}_{\mu}(k)$ is the three possible
polarization vectors, which satisfies the following relations
\cite{itzu}:
\begin{equation}  \varepsilon^{(\lambda)}\cdot k=0,\;\;
\varepsilon^{(\lambda)}\cdot
\varepsilon^{(\lambda')}=\delta_{\lambda \lambda'}, \end{equation}
\begin{equation} \sum_{\lambda =1}^3\varepsilon^{(\lambda)}_{\mu}(k)
\varepsilon^{(\lambda)}_{\nu}(k)=    -(\eta_{\mu
\nu}-\frac{k_{\mu}k_{\nu}}{m^2}) ,\end{equation} and $\eta_{\mu
\nu}=\mbox{diag}(1,-1,-1,-1)$. For simplicity we choose three
five-component vectors $Z^{(\lambda)}$ which obey the transverse
constraints:
$$ Z^{(\lambda)}\cdot \xi=0,$$
\b Z^{(\lambda)}_{\alpha}=(\varepsilon^{(\lambda)}_{\mu}(k),
Z^{(\lambda)}_4=0).\e Here, the de Sitter point $x\equiv x_H(X)$ has
been expressed in terms of the intrinsic cordinate $X^\mu=(X_0=ct,
\vec X)$ measured in units of the de Sitter radius $H^{-1}$:
 $$ x_H(X)=\left(x^0=\frac{\sinh HX^0}{H}, \vec x=\frac{\vec X}{H\parallel
\vec X\parallel} \cosh HX^0 \sin H\parallel \vec X\parallel \right.,$$ \b \left.
x^4= \frac{ \cosh HX^0}{H}  \cos H\parallel \vec X\parallel \right).\label{4.22}\e Note
that $(X^0, \vec X)$ are global coordinates. The compact spherical
nature of spatial part of de Sitter space-time at fixed $X^0$ is apparent in $(\ref{4.22})$.

These solutions are singular at $x\cdot\xi=0,$ and they are not
globally defined due to the ambiguity concerning to the phase
factor. In contrast with the Minkowskian exponentials plan wave,
these waves are singular on three-dimensional light-like manifolds
and can at first be defined only on suitable halves of $X_H$. We
will need an appropriate $i\epsilon$-prescription (indicated below)
to obtain global waves, for detail see \cite{BR}. For a complete
determination, one may consider the solution in the complex the de
Sitter space-time $X_H^{(c)}$ \cite{BROS1,BR}. The complex de Sitter
space-time is defined by
$$ X_H^{(c)}=\left\{ z=x+iy\in  \C^5;\;\;\eta_{\alpha \beta}z^\alpha z^\beta=(z^0)^2-\vec z.\vec z-(z^4)^2=-H^{-2}\right\}$$
\b =\left\{ (x,y)\in  \R^5\times  \R^5;\;\; x^2-y^2=-H^{-2},\;
x.y=0\right\}.\e Let $T^\pm= \R^5+iV^\pm$ be the forward and
backward tubes in $ \C^5$. The domain $V^+$(resp. $V^-)$ stems from
the causal structure on $X_H$: \b V^\pm=\left\{ x\in \R^5;\;\;
x^0\stackrel{>}{<} \sqrt {\parallel \vec x\parallel^2+(x^4)^2}
\right\}.\e The forward and backward tubes of the complex de Sitter
space-time $X_H^{(c)}$ are defined by their respective intersections
with $X_H^{(c)}$, \b {\cal T}^\pm=T^\pm\cap X_H^{(c)}.\e  Finally,
the ``tuboid'' is defined above $X_H^{(c)}\times X_H^{(c)}$ by \b
{\cal T}_{12}=\{ (z,z');\;\; z\in {\cal T}^+,z' \in {\cal T}^- \}.
\e Details are given in \cite{BR}. When $z$ varies in ${\cal T}^+$
(or ${\cal T}^-$) and $\xi$ lies in the positive cone ${\cal C}^+$,
the plane wave solutions are globally defined because the imaginary
part of $(z.\xi)$ has  a fixed sign. The phase is chosen such that
\b \mbox{boundary value of} \; (Hz.\xi)^\sigma \mid_{x.\xi>0}>0.\e
Therefore we have \b \Psi_{1\alpha}(z)={\cal
U}_{\alpha}^{(\lambda)}(z,\xi)(z\cdot \xi)^{-2+i\nu},\e
\begin{equation} \Psi_{2\alpha}(z)={\cal V}_{\alpha}^{(\lambda)}(z,\xi)
(z\cdot \xi)^{-2-i\nu},\e in which $z \in X_H^{(c)} $ and $\xi \in
{\cal C}^+$. The boundary value of the complexified solution is
$$bv (z\cdot \xi)^{-2+i\nu}=(x\cdot \xi)_+^{-2+i\nu}+e^{-i\pi(-2+i\nu)}(x\cdot
\xi)_-^{-2+i\nu},
$$
where $ (x\cdot \xi)_+=\left\{\begin{array}{clcr} 0 & \mbox{for} \;
x\cdot \xi\leq 0\\ (x\cdot \xi) & \mbox{for} \;x\cdot \xi>0. \\
\end{array} \right.$ \cite{geshi}. These solutions are globally defined and they are independent of the
choice of the coordinate system in the de Sitter hyperboloid, i.e.
they are independent of the choice of the metrics of the de Sitter
space.

In the same way as in the Minkowskian space, it is seen that for the
scalar  and vector fields the two solutions are
complex conjugate of each other, but for the spinor
field, there is no such relation between them \cite{BOG}.

\section{The Wightman two-point function}

The Wightman two-point function of spin-$\frac{3}{2}$ field is
defined as \b S^{i \bar
j}_{\alpha\alpha'}(x,x')=<\Omega\mid\Psi^i_\alpha(x){\overline
\Psi}^{\overline i}_{\alpha'}(x')\mid\Omega>,\e where $x,x'\epsilon
X_H$. This function is a solution of equations $(\ref{3.7})$ and
$(\ref{3.10})$ with respect to x and $x'$ respectively. It can be
found in terms of the Wightman two-point function of spinor field,
which was calculated in the previous paper \cite{BOG}.

By using the recurrence formula $(\ref{4.1})$, we define \b
S_{\alpha \alpha'}(x,x')=\theta_{\alpha}. \theta'_{\alpha' }
S_1(x,x')-D_{\frac{3}{2}\alpha}
S_2(x,x'){\gamma^4}\overleftarrow{D}'_{\frac{3}{2}\alpha'}{\gamma^4}
-\gamma^\top_\alpha
S_3(x,x'){\gamma^4}{\gamma'}^\top_{\alpha'}{\gamma^4}.\label{5.2}\e
By imposing the two-point function $S_{\alpha \alpha'}$ to obey
equation $(\ref{3.7})$ and by using the identities of equations
$(\ref{4.2})$-$(\ref{4.5})$, $S_1, S_2$, and $S_3$ must be satisfied
by the following equations: \b
[Q_{\frac{1}{2}}-(\nu^2+\frac{3}{2})]S_1(x,x')=0,\label{5.3} \e
\b[Q_{\frac{1}{2}}-(\nu^2-\frac{3}{2})]S_2(x,x'){\gamma^4}\overleftarrow{D}'_{\frac{3}{2}\alpha'}{\gamma^4}-2(x.\theta')(-S_1(x,x'))=0,\e
\b((Q_1-\frac{7}{2}+\not x\not
\partial^\top )-(\nu^2-\frac{3}{2}))S_3(x,x'){\gamma^4}{\gamma'}^\top_{\alpha'}{\gamma^4}+3\not
x (\theta'.x)S_1(x,x')+(\theta'.\gamma^\top_\alpha)S_1(x,x')=0. \e
By using the conditions $$ x^\alpha S_{\alpha
\alpha'}(x,x')=\gamma^\alpha S_{\alpha
\alpha'}(x,x')=\partial^\alpha S_{\alpha \alpha'}(x,x')=
 \partial^{\top\alpha}S_{\alpha \alpha'}(x,x')=0,$$ we find that
$S_2$ and $S_3$ are given in terms of $S_1$ as
$$-S_2(x,x'){\gamma^4}\overleftarrow{D}'_{\frac{3}{2}\alpha'}{\gamma^4}=\frac{1}{(\nu^2+1)}\left(\frac{2}{3} \theta'_{\alpha'}. \partial^\top +2x.\theta'_{\alpha'}
 +\frac{1}{3}(i\nu+1)\gamma^\top.\theta'_{\alpha'}\not x \right )S_1,$$
 and $$-S_3(x,x'){\gamma^4}{\gamma'}^\top_{\alpha'}{\gamma^4}=-\frac{1}{4}\left(\gamma^\top.\theta'_{\alpha'}\right)S_1+\frac{1}{4}\left(\not\partial^\top+4\not x\right)S_2.$$  The Wightman function can then be written in the form
\begin{equation} S_{\alpha \alpha'}(x,x')=
D_{\alpha \alpha'}(x, \partial^\top ;x', {\partial'}^\top)
S_1(x,x'),\label{5.6}\end{equation} where
$$ D_{\alpha \alpha'}=\theta_\alpha \cdot \theta'_{\alpha'}-\frac{1}{4}\gamma^\top_\alpha\gamma^\top\cdot \theta'_{\alpha'}-\frac{1}{\nu^2+1}\left(\frac{(i\nu-1)}{4}\gamma^\top_\alpha\gamma^\top\cdot \theta'_{\alpha'}+2\partial^\top_\alpha x\cdot\theta'_{\alpha'}\right.$$ $$\left. +\frac{2}{3}\partial^\top_\alpha\theta'_{\alpha'}\cdot\partial^\top -\frac{2}{3}\gamma^\top_\alpha \theta'_{\alpha'} \cdot x \not \partial^\top+ \frac{(i\nu+1)}{3}\partial^\top_\alpha\gamma^\top\cdot\theta'_{\alpha'}\not x-\frac{(i\nu+1)}{3}\gamma^\top_\alpha \theta'_{\alpha'}\cdot x \not x \right.  $$ \b\left. -\frac{1}{6}\gamma^\top_\alpha\theta'_{\alpha'} \cdot\partial^\top \not\partial^\top-\frac{1}{6}i\nu\gamma^\top_\alpha \not x \theta'_{\alpha'} \cdot\partial^\top -\frac{(i\nu+1)}{12}\gamma^\top_\alpha\gamma^\top\cdot\theta'_{\alpha'} \not x \not \partial^\top\right) ,
\e and $S_1$ is solution to $(\ref{5.3})$, which is given by
\cite{BOG} \b S_1=\frac{i}{64\pi}\frac{\nu(1+{\nu^2})}{\sinh (\pi
\nu)}\left[ (2-i\nu)P^{(7)}_{-2-i\nu}(x.x')\not x
-(2+i\nu)P^{(7)}_{-2+i\nu}(x.x')\not x'\right] \gamma^{4}. \e

Similar to the spinor case \cite{BOG}, it is easy to show that this
Wightman two-point function satisfies the following conditions.
\begin{enumerate}
\item[a)] {\bf Positiveness}:
for any spinor-vector test function $f_\alpha \in {\cal D}(X_H)$, we
have
\begin{equation} \int _{X_H \times X_H}  \bar f_i^{\alpha}(x)S_{\alpha
\alpha'}^{ij}(x,x') f_j^{\alpha'}(x')d\sigma(x)d\sigma(x')\geq
0,\end{equation} where $ \bar f$ is the adjoint of $f$ and $d\sigma
(x)$ denotes the de Sitter-invariant measure on $X_H$. ${\cal
D}(X_H)$ is the space of spinor-vector test function $C^\infty$ with
compact support in $X_H$.
\item[b)] {\bf Locality}: for every
space-like separated pair $(x,x')$, {\it i.e.} $x\cdot x'>-H^{-2}$,
\begin{equation} S_{\alpha \alpha'}^{i\overline{j}}(x,x')=-S_{ \alpha'
\alpha}^{\overline{j}i}(x',x),\end{equation}
where $S^{ \bar
j i}_{\alpha'\alpha}(x',x)=<\Omega\mid {\overline
\Psi}^{\overline j}_{\alpha'}(x')\Psi^i_\alpha(x)\mid\Omega>$.
\item[c)] {\bf Covariance}:
\b \Lambda^\alpha_\beta
\Lambda^{\alpha'}_{\beta'}g^{-1}S_{\alpha\alpha'}\bigl(\Lambda(g)x,\Lambda(g)x'\bigr)i(g)=S_{\beta\beta'}(x,x'),
\label{covarianza} \e where  $\Lambda \in SO_0(1,4)$ , $g\in
Sp(2,2)$ and $ g\gamma^\alpha g^{-1}=\Lambda_\beta^\alpha
\gamma^\beta$. $i(g)$ is the group involution defined by
\begin{equation}
i(g)=-\gamma^{4}g\gamma^{4}. \label{invol}
\end{equation}
\item[d)] {\bf Transversality}:
\begin{equation} x\cdot S(x,x')=0=x'\cdot S(x,x'),\end{equation}
\item[e)] {\bf Divergencelessness}:
\begin{equation} \partial_x\cdot S(x,x')=0=\partial_{x'}\cdot
S(x,x'),\end{equation}
\item[f)] {\bf Normal analyticity}:
$S_{\alpha \alpha' }(x,x')$ is the boundary value (in the distributional sense ) of
an analytic function ${\cal S}_{\alpha \alpha'}(z,z')$.
\end{enumerate}
${\cal S}_{\alpha \alpha'}(z,z')$ is maximally analytic, i.e., can be analytically continued to the ``cut domain''\cite{BR,BOG}:
$$\Delta=\{ (z,z') \in X_H^{(c)} \times X_H^{(c)} \;\; :\;\; (z-z')^2\leq 0 \}. $$
The Wightman two-point function $S_{\alpha \alpha' }(x,x')$ is the
boundary value of ${\cal S}_{\alpha \alpha'}(z,z')$ from ${\cal
T}_{12}$ and the ``permuted Wightman function'' $S_{\alpha' \alpha
}(x',x)$ is the boundary value of ${\cal S}_{\alpha \alpha'}(z,z')$
from the domain:
$$ {\cal T}_{21}=\{ (z,z');\;\; z'\in {\cal T}^+,z \in {\cal T}^- \}. $$

\section{Quantum field}

The existence of a two-point function with the above-mentioned
properties allows us to define (via the reconstruction theorem
\cite{[SW]}) a massive spin-$\frac{3}{2}$ field operator
$\Psi_{\alpha}$, satisfying the field equation, as an
operator-valued distribution on $X_H$ defined on (a dense domain in)
a separable Hilbert space ${\cal H}$. The Hilbert space ${\cal H}$
can be described as the Hilbertian sum \cite{[N.B]}:
\begin{equation}
{\cal H}={\cal H}_{0}\oplus\left[\oplus^{\infty}_{n=1}A(\otimes
^{n}{\cal H}_{1})\right],
\end{equation}
where $A$ denotes the antisymmetrization operation and
\begin{equation}
{\cal H}_{0}=\{\lambda\Omega, \lambda\in\C \}.
\end{equation}
${\cal H}_{1}$ is defined as follows (given the positive
definite inner product):
\begin{equation}
\langle h_,f\rangle =\int_{X_H\times
X_H}\overline{h}(x_{1})S(x_{1},x_{2})f(x_{2})d\sigma(x_{1})d\sigma(x_{2}).\label{innprod}
\end{equation}
A regular element $h_\alpha^i\in {\cal H}_{1}$ is a class of
spinorial-vector test functions $h(x)$ (i.e. in ${\cal D}(X_H)$)
modulo the functions g such that the corresponding seminorm
vanishes. The full Hilbert space ${\cal H}_{1}$ is the completion
with this norm of the space of regular elements. In terms of
creation and annihilation operators the smeared field operators
$\Psi_\alpha(f)$ are realized as
\begin{eqnarray}
&&\bigl( \Psi(f)h\bigr)^{(n)}(x_{1},i_{1},\alpha_1;x_{2},i_{2},\alpha_2;\dots
;x_{n},i_{n},\alpha_n) =\nonumber\\
&&{\sqrt{n+1}}\int_{X_H\times
X_H}f_\alpha^{i}(x)(S^{\alpha\beta})_{i \overline{j}}(x,y)
h^{(n+1)}(y,\overline{j},\beta;x_{1},i_{1},\alpha_n;\dots
;x_{n},i_{n},\alpha_n)d\sigma(x)d\sigma(y) \nonumber\\ &&+{1\over
{\sqrt
n}}\sum_{k=1}^{n}(-1)^{k+1}f_{\alpha_k}^{i_{k}}(x_{k})h^{(n-1)}(x_{1},i_{1},\alpha_1;\dots
;{\hat x}_{k},{\hat i}_{k},{\hat \alpha}_k;\dots
;x_{n},i_{n},\alpha_n).\label{smfield}
\end{eqnarray}

The field operator $\Psi_{\alpha}$ in terms of plane waves and
creation and annihilation operators reads as
$$\Psi_\alpha(x)=\int_{\Gamma}\sum_{a,\lambda}\left(c_{a\lambda}(\xi){\cal U}^{a\lambda}_\alpha(x,\xi)\left[(Hx\cdot \xi)_+^{-2+i\nu}+e^{-i\pi(-2+i\nu)}(Hx\cdot
\xi)_-^{-2+i\nu}\right]\right.$$
$$\left.+d^{\dag}_{a\lambda}(\xi){\cal
V}^{a\lambda}_\alpha(x,\xi)\left[(Hx\cdot
\xi)_+^{-2+i\nu}+e^{-i\pi(-2+i\nu)}(Hx\cdot
\xi)_-^{-2+i\nu}\right]\right)d\mu_\Gamma(\xi),$$
\begin{equation}
\equiv \Psi_\alpha^{(-)}(x)+\Psi_\alpha^{(+)}(x). \label{devf}
\end{equation}
By using this field operator, we have the two-point function $(5.6)$ for any given value of the
mass parameter $\nu$. Here $\Gamma$ denotes an orbital basis of
${\cal C}^+$. $d\mu_\Gamma(\xi)$ is an invariant measure defined by
\cite{BR} \b d\mu_\Gamma(\xi)=i_\Xi w_{{\cal C}^+}\mid_\Gamma,\e
where $i_\Xi w_{{\cal C}^+}$ denotes the $3$-form on ${\cal C}^+$
obtained from the contraction of the vector field $\Xi$ with the
volume form \b w_{{\cal C}^+}=\frac{d\xi^0\wedge \cdots \wedge
d\xi^4 }{d(\xi\cdot \xi)}.\e The operators $c$ and $d$ annihilate
the fundamental state and  $c^\dag$ and $d^\dag$ create ``one
particle'' states:$$ c_{a\lambda}(\xi,\nu)|\Omega>=0,
\;\;d_{a\lambda}(\xi,\nu)|\Omega>=0,$$
$$ c_{a\lambda}^{\dag}(\xi,\nu)|\Omega>=|(\xi,\lambda,a)^{(c)}>,
\;\;d_{a\lambda}^{\dag}(\xi,\nu)|\Omega>=|(\xi,\lambda,a)^{(d)}>.$$
By using the above conditions one can show that $\Psi_\alpha(x)$
satisfies the locality properties \cite{BOG}: $$
\bigl\{\Psi^{i}_\alpha(x_{1}),
\overline{\Psi}^{\overline{j}}_\beta(x_{2})\bigr\}=0,
\label{anticomm}
$$ for every space-like separated pair $(x_{1},x_{2})$ in $X_H$.

\section{ Conclusions}

In this paper, we have considered the ``massive'' field
associated to the principal series of the de Sitter group $SO_0(1,4)$ with
$<Q_\nu>=\nu^2-\frac{3}{2},\;\;\nu > \frac{3}{2}$. For the discrete  series
$$<Q_{\frac{3}{2}}>=-\frac{15}{4}-(q+1)(q-2),\;\;q=\frac{1}{2},\;\; \frac{3}{2},$$ we can replace $\nu$
by $\nu=0$ and  $\nu=\pm i$ for $q=\frac{1}{2},\;\; \frac{3}{2},$
respectively. For the discrete series only the representations
$\Pi^{\pm}_{\frac{3}{2},\frac{3}{2}}$ have a physically meaningful
Poincar\'e limit. These are precisely the ``massless'' spinor-vector
field and $\nu$ must be replaced by $\pm i$ in the previous
formulas. But the solutions of equations $(\ref{4.19})$ and
$(\ref{4.20})$ are divergent at this limit ($\nu=\pm i$). This type
of singularity is actually due to the auxiliary conditions
($\partial \cdot \Psi= \gamma \cdot\Psi=0$) imposed in order to
associate this field with a specific unitary irreducible
representation of the de Sitter group. It can be as well understood
from equation (\ref{4.11}), allowing one to determine $\psi_2$ and
$\psi_3$ in terms of $\psi_1$. To solve this problem, the auxiliary
conditions must be dropped out and then the field equation becomes
gauge invariant, which had been studied in Riemannian space-time in
the intrinsic coordinate \cite{RED'KOV}. This situation will be
considered in the ambient space notation and from a group
theoretical point of view in a forthcoming paper \cite{taaz}.

\vspace{3mm} \noindent {\bf{Acknowledgments}}: We would like to
thank S. Rouhani, M. R. Tanhai and T. Parvizi for their interest in
this work. We also thank the referees for their useful comments and
suggestions.

\setcounter{equation}{0}
\begin{appendix}

\section{Appendix}
\catcode `\@=11 \@addtoreset{equation}{section}
\def\theequation{\Alph{section}.\arabic{equation}}
\catcode `\@=12 \catcode `\@=11 \@addtoreset{proposition}{section}
\def\theproposition{\arabic{section}.\arabic{proposition}}
\catcode `\@=12 \catcode `\@=11 \@addtoreset{theorem}{section}
\def\thetheorem{\arabic{section}.\arabic{theorem}}
\catcode `\@=12 \catcode `\@=11
\label{A}

\subsection{The solution of the first order field equation}

In this appendix, we would like to find the solution of the first
order equation $(\ref{3.9})$. By substituting $\Psi_\alpha(x)$,
which is given by equation $(\ref{4.1})$, we have \b \left(-i\not
x\not\partial^\top+2i+\nu\right)\left( Z^\top_\alpha
\psi_1+D_{\frac{3}{2}\alpha}
\psi_2+\gamma^\top_\alpha\psi_3\right)=0.\e By using the following
identities: \b \not x\not\partial^\top
Z_\alpha^\top\psi_1=-\gamma_\alpha^\top\not xx.Z\psi_1+x_\alpha\not
x\not Z^\top\psi_1+Z_\alpha^\top\not x\not\partial^\top\psi_1,\e \b
\not x\not\partial^\top\partial^\top_\alpha\psi_2=\not
x\partial^\top_\alpha\not\partial^\top\psi_2+x_\alpha\not
x\not\partial^\top\psi_2+\partial^\top_\alpha\psi_2,\e \b \not
x\not\partial^\top\gamma^\top_\alpha\not
x\psi_2=5\gamma^\top_\alpha\not
x\psi_2-4x_\alpha+\gamma^\top_\alpha\not\partial^\top\psi_2+2\not
x\partial^\top_\alpha\not x\psi_2, \e \b \not
x\not\partial^\top\gamma^\top_\alpha\psi_3=
\gamma^\top_\alpha\psi_3+4x_\alpha\not x\psi_3+2\not
x\partial^\top_\alpha\psi_3+\gamma^\top_\alpha\not
x\not\partial^\top\psi_3,\e and by putting equation $(4.10)$ in
equation $(A.5)$ and using of equations $(A.2-4)$ we have \b (\not
x\not\partial^{\top}-2+i\upsilon)\psi_1=0,\e
$$ (i\upsilon-1)\not
x\psi_2+\frac{i\upsilon-1}{4}\not\partial^{\top}\psi_2+\frac{1}{4}\not
x\not\partial^{\top}\not\partial^{\top}\psi_2$$\b=\not
xx.Z\psi_1-\frac{1}{4}\not Z^{\top}\psi_1+\frac{1}{4}\not
x\not\partial^{\top}\not Z^{\top}\psi_1+\frac{i\upsilon}{4}\not
Z^{\top}\psi_1\e \b \not\partial^{\top}\psi_2=2(i\upsilon-1)\not
x\psi_2-\not Z^{\top}\psi_1.\e By substituting equation $(A.8)$ in
equation $(A.7)$, $\psi_2$ and $\psi_3$ are obtained as follows:
 \b
\psi_2=\frac{1}{(\nu^2+1)}\left(2x.Z+\frac{2}{3}Z. \partial^\top
+\frac{1}{3}(i\nu+1)\not Z^\top\not x
 \right)\psi_1,\e
  \b \psi_3=-\frac{1}{4}\left[\gamma. Z^\top -\frac{\gamma.
\partial^\top +4\not x}{(\nu^2+1)}\left(2x.Z+\frac{2}{3}Z. \partial^\top
+\frac{1}{3}(i\nu+1)\not Z^\top\not x
 \right)\right]\psi_1.\e

\subsection{Spinor-vector ${\cal U}$ and ${\cal V}$}

By taking the derivative of plane wave,
$$ \bar{\partial}_\alpha(Hx.\xi)^\sigma=\sigma \bar{\xi}_\alpha (Hx.\xi)^{\sigma-1},$$
we obtain
$$ {\cal V}_\alpha(x,\xi,Z)=Z^\top_{\alpha}-\frac{1}{4}\gamma^\top_\alpha\not Z^\top-\frac{1}{\nu^2+1}
\left[ \frac{i\nu+3}{3}Z^\top_{\alpha}\right. $$
$$+\xi^\top_\alpha\left(\frac{(8i\nu-16)}{3}\frac{x.Z}{x.\xi}+\frac{(i\nu-1)(i\nu-2)}{3}\frac{\not Z^\top \not x}{x.\xi}+
\frac{2(i\nu-2)(i\nu-3)}{3}\frac{Z.\xi^\top}{(x.\xi)^2}\right)$$
\b\left. +\gamma^\top_\alpha\left(-\frac{(2i\nu+4)}{3}x.Z\not
x-\frac{(i\nu+1)(i\nu-2)}{3}\frac{\not x Z.\xi^\top }{x.\xi}
 -\frac{(\nu^2-4i\nu+5)}{12}\not Z^\top\right)\right]
,\e

$$ {\cal U}_\alpha(x,\xi,Z)=Z^\top_{\alpha}-\frac{1}{4}\gamma^\top_\alpha\not Z^\top-\frac{1}{\nu^2+1}
\left[ \frac{-i\nu+3}{3}Z^\top_{\alpha}\right. $$
$$+\xi^\top_\alpha\left(-\frac{(8i\nu+16)}{3}\frac{x.Z}{x.\xi}-\frac{(i\nu+1)(i\nu+2)}{3}\frac{\not Z^\top \not x}{x.\xi}+
\frac{2(i\nu+2)(i\nu+3)}{3}\frac{Z.\xi^\top}{(x.\xi)^2}\right)$$
\b\left. +\gamma^\top_\alpha\left(\frac{(-2i\nu+4)}{3}x.Z\not
x+\frac{(i\nu-1)(i\nu+2)}{3}\frac{\not x Z.\xi^\top }{x.\xi}
 -\frac{(\nu^2+4i\nu+5)}{12}\not Z^\top\right)\right]
.\e

\subsection{Two-point function}

Here, the two-point function is calculated with respect to $x'$,
which satisfies equation $(\ref{3.10})$. Putting equation
$(\ref{5.2})$ in equation $(\ref{3.10})$, we obtain \b
S_1(x,x'){\gamma^4}\left[\overleftarrow{Q'}_{\frac{1}{2}}-(\nu^2+\frac{3}{2})\right]{\gamma^4}=0,
\e \b
D_{\frac{3}{2}\alpha}S_2(x,x'){\gamma^4}\left[\overleftarrow{Q'}_{\frac{1}{2}}-(\nu^2-\frac{3}{2})\right]{\gamma^4}-2S_1(x,x'){\gamma^4}(x'.\theta){\gamma^4}=0,\e
 $$\gamma^\top_\alpha S_3(x,x'){\gamma^4}\left((\overleftarrow{Q'}_1-\frac{7}{2}+\overleftarrow{\not\partial}{'^\top}\not x)-(\nu^2-\frac{3}{2})\right){\gamma^4}$$\b-3S_1(x,x'){\gamma^4}\not x' (\theta.x'){\gamma^4}-S_1(x,x'){\gamma^4}\left(\theta.{\gamma'}^\top_{\alpha'}\right){\gamma^4}=0.
\e By using the subsidiary conditions: $$ x'^{\alpha'}S_{\alpha
\alpha'}(x,x')=S_{\alpha
\alpha'}(x,x'){\gamma^4}\gamma^{\alpha'}{\gamma^4} =S_{\alpha
\alpha'}(x,x')\overleftarrow{\partial}^{\alpha'}= S_{\alpha
\alpha'}(x,x')\overleftarrow{\partial}^{\top\alpha'}=0,$$ we can
write $S_2$ and $S_3$ in terms of $S_1$ :
$$D_{\frac{3}{2}\alpha}S_2(x,x')=\frac{1}{(\nu^2+1)}S_1\left(2x'\cdot\theta_\alpha+\frac{2}{3}\theta_\alpha.\overleftarrow{\partial}{'^\top}-\frac{1}{3}(i\nu+1){\gamma^4}\gamma'^\top\cdot\theta_\alpha\not x'{\gamma^4}
\right),$$ and $$\gamma^\top_\alpha
S_3(x,x')=-\frac{1}{4}S_1\gamma^4\left(\gamma^{'\top}.\theta_{\alpha}\right)\gamma^4-\frac{1}{4}S_2\gamma^4\left(\overleftarrow{\not\partial}{'^\top}+4\not
x'\right)\gamma^4.$$ Finally the two-point function obtain similar
to equation $(\ref{5.6})$ as
\begin{equation} S_{\alpha \alpha'}(x,x')=
S_1(x,x')D'_{\alpha \alpha'}(x, \overleftarrow{\partial}{'^\top}
;x',\overleftarrow{\partial}{'^\top}),\end{equation} where
$$ D'_{\alpha' \alpha}=\theta'_{\alpha'} \cdot \theta_{\alpha}-\frac{1}{4}\gamma^4\gamma'^\top_{\alpha'}\gamma'^\top\cdot \theta_{\alpha}\gamma^4-\frac{1}{\nu^2+1}\left(\frac{(i\nu-1)\gamma^4\gamma'^\top_{\alpha'}\gamma'^\top\cdot \theta_{\alpha}\gamma^4}{4}+2x'\cdot\theta_{\alpha}\overleftarrow{\partial}{'^\top}_{\alpha'}\right.$$ $$ \left.+\frac{2
\overleftarrow{\partial}{'^\top}\cdot\theta_{\alpha}\overleftarrow{\partial}{'^\top}_{\alpha'}}{3}
+\frac{2\overleftarrow{\not\partial}{'^\top}\gamma'^\top_{\alpha'}
\theta_{\alpha} \cdot x'}{3}+
\frac{i\nu+1}{3}\gamma'^\top\cdot\theta_{\alpha}\not
x'\overleftarrow{\partial}{'^\top}_{\alpha'}-\frac{i\nu+1}{3}\gamma^4\gamma'^\top_{\alpha'}
\theta_{\alpha}\cdot x' \not x'\gamma^4\right.
$$
\b\left.+\frac{\overleftarrow{\not\partial}{'^\top}\overleftarrow{\partial}{'^\top}\cdot\theta_{\alpha}\gamma'^\top_{\alpha'}}{6}
-\frac{i\nu \overleftarrow{\partial}{'^\top}\cdot\theta_{\alpha}
\gamma'^\top_{\alpha'} \not x'}{6} +\frac{i\nu+1}{12}
\overleftarrow{\not\partial}{'^\top}\gamma'^\top_{\alpha'}\gamma'^\top\cdot\theta_{\alpha}
\not x'\right). \e

 \end{appendix}

\end{document}